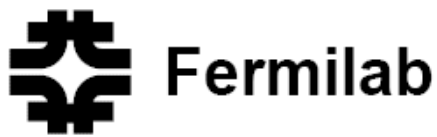

# LONG BASELINE NEUTRINO EXPERIMENT TARGET MATERIAL RADIATION DAMAGE STUDIES USING ENERGETIC PROTONS OF THE BROOKHAVEN LINEAR ISOTOPE PRODUCTION (BLIP) FACILITY [*†]

N. Simos#; J. O Conor; BNL, Upton, NY 11973, USA,
P. Hurh, N. Mokhov; FNAL, Batavia, IL 60510, USA
Z. Kotsina, Democritos National Center for Scientific Research, Greece

## Abstract

One of the future multi-MW accelerators is the LBNE Experiment where Fermilab aims to produce a beam of neutrinos with a 2.3 MW proton beam as part of a suite of experiments associated with Project X. Specifically, the LBNE Neutrino Beam Facility aims for a 2+ MW, 60-120 GeV pulsed, high intensity proton beam produced in the Project X accelerator intercepted by a low Z solid target to facilitate the production of low energy neutrinos. The multi-MW level LBNE proton beam will be characterized by intensities of the order of 1.6 e+14 p/pulse, σ radius of 1.5-3.5 mm and a 9.8 μs pulse length. These parameters are expected to push many target materials to their limit thus making the target design very challenging. To address a host of critical design issues revealed by recent high intensity beam on target experience a series of experimental studies on radiation damage and thermal shock response conducted at BNL focusing on low-Z materials have been undertaken with the latest one focusing on LBNE.

---

[*]Work supported by Fermi Research Alliance, LLC under contract No. DE-AC02-07CH11359 with the U.S. Department of Energy.
[†]Presented paper at the 52nd ICFA Advanced Beam Dynamics Workshop on High-Intensity and High-Brightness Hadron Beams, HB2012, Beijing, September 17-21, 2012.
[#]simos@bnl.gov

# LONG BASELINE NEUTRINO EXPERIMENT TARGET MATERIAL RADIATION DAMAGE STUDIES USING ENERGETIC PROTONS OF THE BROOKHAVEN LINEAR ISOTOPE PRODUCTION (BLIP) FACILITY*

N. Simos#; J. O Conor; BNL, Upton, NY 11973, USA,
P. Hurh, N. Mokhov; FNAL, Batavia, IL 60510, USA
Z. Kotsina, Democritos National Center for Scientific Research, Greece

*Abstract*

One of the future multi-MW accelerators is the LBNE Experiment where Fermilab aims to produce a beam of neutrinos with a 2.3 MW proton beam as part of a suite of experiments associated with Project X. Specifically, the LBNE Neutrino Beam Facility aims for a 2+ MW, 60-120 GeV pulsed, high intensity proton beam produced in the Project X accelerator intercepted by a low Z solid target to facilitate the production of low energy neutrinos. The multi-MW level LBNE proton beam will be characterized by intensities of the order of 1.6 e+14 p/pulse, σ radius of 1.5-3.5 mm and a 9.8 µs pulse length. These parameters are expected to push many target materials to their limit thus making the target design very challenging. To address a host of critical design issues revealed by recent high intensity beam on target experience a series of experimental studies on radiation damage and thermal shock response conducted at BNL focusing on low-Z materials have been undertaken with the latest one focusing on LBNE.

## INTRODUCTION

High-performance targets under consideration to intercept multi-MW proton beams of a number of new particle accelerator initiatives depend almost entirely on the ability of the selected materials to withstand both the induced thermo-mechanical shock and simultaneously resist accumulated dose-induced damage which manifests itself as changes in material physio-mechanical properties. One of the future multi-MW accelerators is the LBNE Experiment where Fermilab plans to produce a beam of neutrinos with a 2.3 MW proton beam as part of a suite of experiments associated with Project X. Specifically, the LBNE Neutrino Beam Facility aims for a 2+ MW, 60-120 GeV pulsed, high intensity proton beam produced in the Project X accelerator intercepted by a low Z solid target to facilitate the production of low energy neutrinos. The multi-MW level LBNE proton beam will be characterized by intensities of the order of 1.6 e+14 p/pulse, σ radius of 1.5-3.5 mm and a 9.8 µs pulse length. These parameters are expected to push many target materials to their limit thus making the target design very challenging. Recent experience from operating high intensity beams on targets have indicated that several critical design issues exist namely thermal shock, heat removal, radiation damage, radiation accelerated corrosion effects, and residual radiation within the target envelope. A series of experimental studies on radiation damage and thermal shock response conducted at BNL and focusing on low-Z materials have unraveled potential issues regarding the damageability from energetic particle beams which may differ significantly from thermal reactor experience.

To address irradiation damage from energetic particles proton a wide array of materials considered to support high power experiments have been studied extensively using the BNL 200 MeV proton beam of the Linac and utilizing the target station of the Linear Isotope Producer (BLIP) where 20-24 kW of proton beam power (~110 µA current) are effectively used to irradiate target materials under consideration. Of interest to the LBNE operating with 120 GeV protons are low Z target materials and their operational life that is expected to be limited by irradiation damage. Instead of extrapolating from thermal neutron damage experience accumulated in fission reactors damage from the energetic protons at BNL Linac was sought to deduce target lifetime estimates. Based on first principles of energetic particle interaction with matter, it is anticipated that the damage to these sought low Z materials will be greater at these MeV proton energy levels than the 120 GeV of the LBNE beam.

Extensive simulations performed by MARS15 indeed revealed that for the NuMI/LBNE experiment operating at 120 GeV with beam σ = 1.1mm and 4.0e20 protons/year is expected to see peak damage in graphite of ~0.45 dpa while for the BLIP configuration with a beam energy of 165 MeV and σ = 4.23mm and 1.124e22 protons on target/year the expected damage of 1.5 dpa. Guided by these analytical predictions on carbon materials of interest the equivalent damage of LBNE operations at the 700 kW level may be achieved with ~7-8 weeks irradiation using the BNL 181 MeV Linac proton beam and ~5mm sigma.

The main objectives of the irradiation experiment were to (a) assess the effect that the operating environment has on the onset and acceleration of structural degradation of graphite and carbon composites when exceeding certain fluence levels of energetic protons that are far lower than the thermal neutron fluences, (b) the role of irradiation temperature in restoring damage induced by the irradiating beam and (c) the variability in damage experienced by different grades of graphite with distinct polycrystalline structure as well as other materials such as carbon and h-BN that exhibit lattice similarities.

*Work supported by Fermi Research Alliance, LLC under contract No. DE-AC02-07CH11359 with the U.S. Department of Energy.
#simos@bnl.gov

Discussed in the paper are (a) the reasoning for selecting the test array of materials to be irradiated at the BNL isotope producer facility, (b) the special test conditions including the irradiation temperatures, (c) the effects of irradiation on dimensional changes and annealing of damage as seen in post-irradiation analysis including a discussion on the theoretical aspects of the process and (d) effects on mechanical behavior seen by stress-strain tests as well as elastic modulus changes supplemented by ultrasonic-based evaluation.

## LBNE TARGET STUDIES AT BNL BLIP

Graphite is the primary candidate for the LBNE Experiment and of interest are its grade varieties of which have been used extensively as primary targets and beam collimators. In addition, carbon-carbon composites and the low-Z alloy of beryllium with aluminum (AlBeMet) are possible alternatives. Carbon-carbon composites, in particular, which appear in two- and three-dimensional weave structure exhibit very low thermal expansion along the carbon fiber orientation and are much stronger than graphite. These two attributes directly influencing shock absorbing capabilities led to their consideration for higher power targets than what graphite have served to-date and for beam intercept elements in the LHC collimators. BNL studies using the 24 GeV AGS beam with a tightly focused proton pulse (0.3mm x 0.9mm rms and $4 \times 10^{12}$ 24 GeV protons) confirmed the superiority of carbon composite structures in mitigating thermal shock. Irradiation with energetic protons, however to fluences of ~$10^{21}$ p/cm$^2$ in direct contact with cooling water showed a serious degradation of the C/C structure. The concern over the operating environment, proton energy and flux effects have prompted the further studies of the carbon composites under the LBNE Experiment initiative. To address the energy and flux effects specific to the study MARS15 studies were conducted revealing anticipated damage. To address the environment effects on structural degradation the experiment considered both water and argon environments. Shown in Fig. 1 are SEM images of the carbon fiber structure indicating significant porosity that may influence its behavior while irradiated. Table 1 lists the final test material array along with the application and the relevant issues that have been experienced.

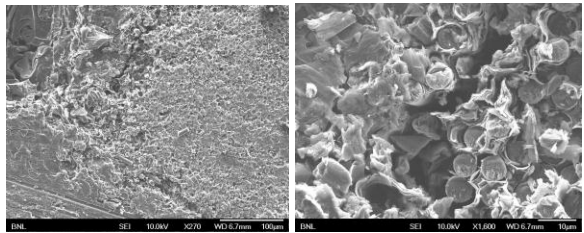

Figure 1: Micrographs of C/C composite fiber structure

Table 1: BLIP Material Matrix

| Material | Motivation |
|---|---|
| C-C Composite (3D) | Observed damage at low dose |
| POCO ZXF-5Q | NuMI/NOvA target material |
| Toyo-Tanso IG-430 | Nuclear grade for T2K |
| Carbone-Lorraine 2020 | CNGS target material |
| SGL R7650 | NuMI/NOvA baffle material |
| St.-Gobain AX05 h-BN | Hexagonal Boron Nitride |

Interest in graphite was prompted by the fact that graphite irradiated with 200 MeV protons alongside with C/C composite and with direct water cooling also exhibited accelerated structural integrity loss at similar fluences. Further because the graphite lattice exhibits anisotropic dimensional changes under irradiation which will affect the polycrystalline structure differently depending on the grade and its formation understanding the behavior of the various grades under irradiation and annealing is valuable information. Two-dimensional C/C fiber composite structures that have been studied in the recent past as LHC collimators resemble the graphite lattice behavior but in the global sense.

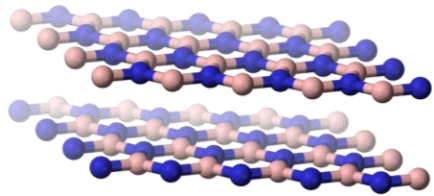

Figure 2: Lattice structure of hexagonal boron nitride

The interest in h-BN whose lattice structure is shown in Fig. 2 is because it is a close neighbor to graphite with most stable hexagonal crystalline form. Within each layer, boron and nitrogen atoms are bound by strong covalent bonds whereas the layers are held together by weak van der Waals forces (similarly for graphite layers in its lattice structure). Significant differences exist in the interlayer "communication" between h-BN and graphite due to the fact that boron atoms lying over and above nitrogen atoms in two adjacent layers affecting the polarity of the B-N bonds. Still, h-BN and graphite are very close neighbors. Due to excellent thermal and chemical stability, boron nitride ceramics enjoy high-temperature applications.

*Experiment Irradiation Temperatures*

While the temperature during which irradiation takes place is not affecting the defect production it is a crucial parameter in annealing the defects that are generated and do not recombine during the initial, fast process by inducing mobilization of interstitials and vacancies based on activation energies. Theoretical aspects of defect mobilization and its influence on the physio-mechanical properties will be discussed in more detail in a later section. The underlying principle, however, of interstitial atom movement up to a given activation energy or temperature has been utilized in post-irradiation analysis to pin-point the temperature in the tested samples during proton irradiation at the BNL target station. The principle is shown in Fig. 3 where the irradiation temperature is identified as the point of deviation between the unirradiated and irradiated dimensional change curves of graphite (similarly with C/C composites). During low-temperature irradiation the interstitial atoms knocked from the basal planes that require as activation energy the one associated with the irradiation temperature have been prompted to return to vacant positions on the planes and therefore the unirradiated and irradiated volumetric change in the sample are identical to this threshold. High fidelity models of thermal analysis utilized prior to the test have predicted similar irradiation temperatures. Peak irradiation temperatures in argon encapsulated samples of 160-180 $^{o}$C were experienced with slightly lower temperatures in water cooled C/C samples (~140 $^{o}$C).

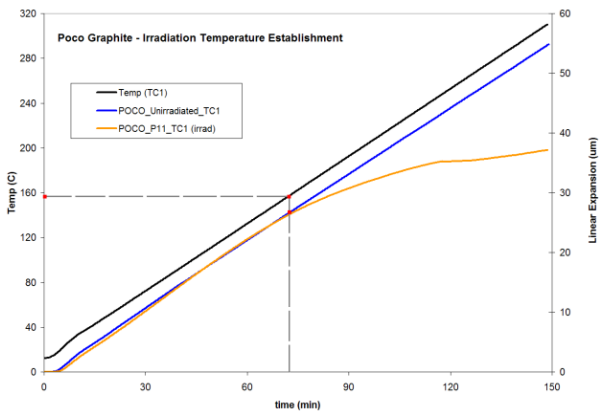

Figure 3: Post-irradiation assessment of temperature

*Dimensional Change and Annealing*

Anisotropy and dimensional change in graphite is the result of interstitial and vacancy formation due to energetic neutrons or protons impinging on the lattice and the coalescence of interstitials into molecules between basal planes and the collapse of vacancies formed on the basal plane. These result in crystalline dimensional change manifested as growth perpendicular to the planes and shrinkage parallel to the planes. While the operating temperature has no effect on the rate of formation and the immediate annealing of interstitials, it can control the activation energy of stable interstitials and vacancies making them mobile and thus reversing the volumetric change that has occurred and in the process partially or fully restoring some of the physical and mechanical properties [3]. Therefore post-irradiation analyses focusing on the dimensional restoration or thermal expansion can reveal key information on the damage induced by the irradiating species.

Post irradiation measurements on graphite, carbon fiber composites and h-BN are shown in Figures 4-11. Figure 4 establishes the baseline (unirradiated state) of graphites and 2D C/C composite in the direction normal to the fiber planes (resembling graphite crystal lattice with interstitial atoms between the planes). The dimensional change following irradiation is depicted in Fig. 5 where a common trend is being observed. Specifically, shrinkage is observed at temperatures above the irradiation temperature which is the result of interstitial atom mobilization which leave their inter-plane position to return to vacant positions on the plane leading to the restoration of the pre-irradiation volume. Recall that because of the accumulation of interstitial atoms between the basal planes the graphite crystal increased in volume along the direction normal to the planes. In a subsequent thermal cycle, shown in Fig. 6 the linear dimensional change up to the annealing temperature of the previous cycle has been restored.

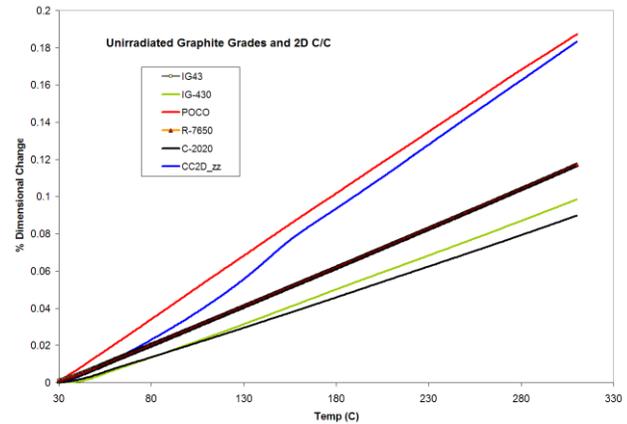

Figure 4: Unirradiated dimensional change of graphite

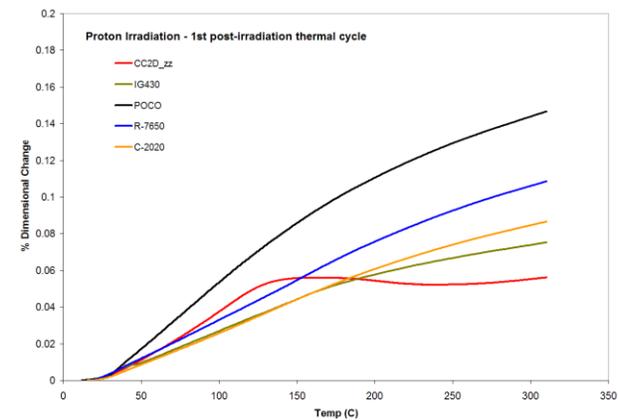

Figure 5: Dimensional change in irradiated graphites

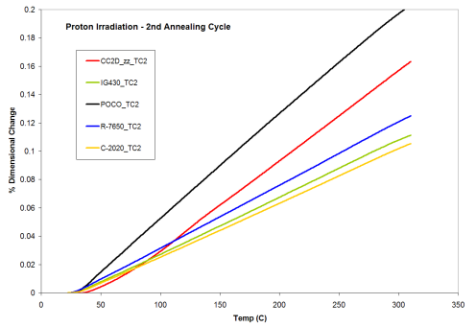

Figure 6: Achieved annealing of irradiated graphite

The activation energy or annealing temperature that mobilizes interstitials at low temperatures and vacancies at higher temperatures is further demonstrated in Fig. 7 in the annealing of irradiated IG-430 graphite and compared with the unirradiated counterpart. While the first pass and dwell at 200 $^{o}$C clearly shows annealing in the irradiated IG-430 (slopping of the dimensional change) which is also evident at a higher temperature (300 $^{o}$C), subsequent dwell at 200 $^{o}$C shows that all interstitial atoms requiring activation energy associated with 200 $^{o}$C have been mobilized and returned to the lattice planes.

The effect of irradiation temperature coupled with the cooling environment of 3D C/C composite is shown in Fig. 8. C/C composite in argon atmosphere which implies higher irradiation temperature and its annealing are compared with water cooled C/C. As shown, more unrecoverable damage is associated with the water-cooled C/C composite. Due to the similarity in the dimensional change behavior between h-BN and C/C along the fibers, the annealing of irradiated h-BN is compared with an unirradiated sample in Fig. 9.

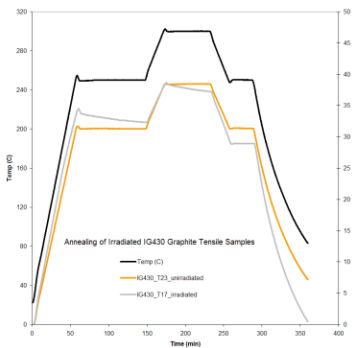

Figure 7: Annealing of irradiated IG-430 graphite

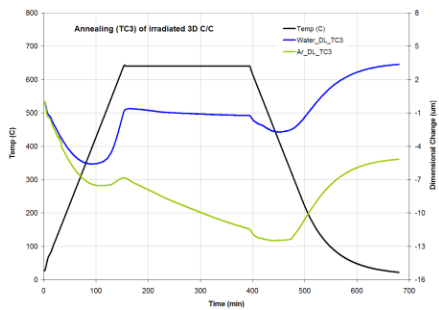

Figure 8: Annealing in 3D C/C (argon vs. water)

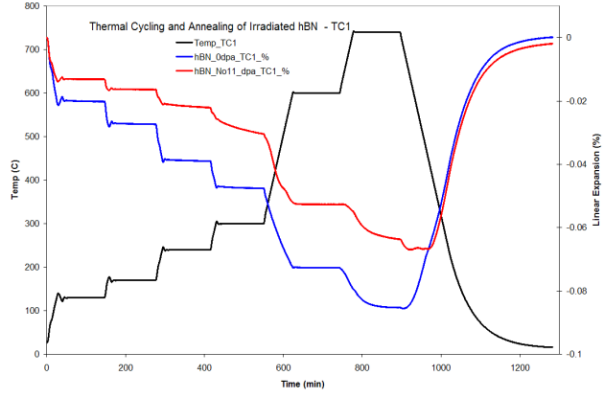

Figure 9: Irradiated h-BN annealing process

The annealing of irradiated graphite IG-43 at higher temperatures where vacancies are expected to become mobile is shown in Fig. 11.

## Mechanical Behavior

The effects of irradiation as well as of post-irradiation annealing on the mechanical properties of the various graphite grades were assessed through stress-strain testing using a Tinius-Olsen mechanical tester. Of interest, in addition to the assessment of strength and modulus changes were the effect that annealing to higher than the irradiation temperature has on the mechanical response as a result of damage reversal due to mobilization of interstitials and vacancies formed during irradiation. Figures 12-14 depict stress-strain behavior for irradiated IG-430, POCO and R7650 graphites and compares them with the unirradiated state. Also shown is the effect of annealing prior to the mechanical testing of the specimens. Worth noting is that irradiation not only increases the ultimate strength of all graphites (typical in all materials by impeding dislocation or crystal plane slide due to the agglomeration of interstitials) but also significantly changes the elastic modulus of graphite something that does not occur in metallic lattices.

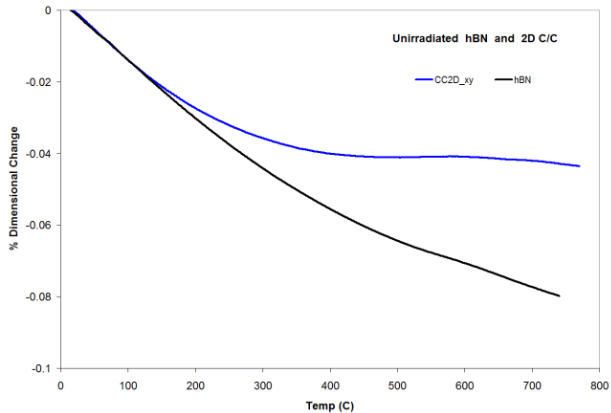

Figure 10: h-BN and fiber orientation of C/C

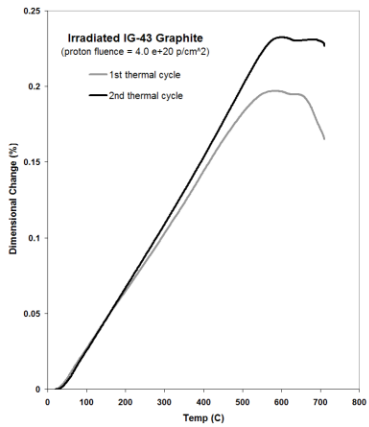

Figure 11: High temperature annealing of graphite IG-43

With an integration of annealing, volumetric change, density change and ultrasonic velocity measurements the effect of irradiation on graphite modulus has been studied in parallel with the changes observed in the stress-strain curves. Similar studies on H-451 graphite have been conducted and reported in [3] where estimates of elastic modulus E change were deduced as a function of neutron fluence. The significant increase in E observed in the LBNE irradiation studies was also evident in [3]. Fig. 15 depicts the annealing process and the measured sound velocity change in irradiated C-2020 graphite.

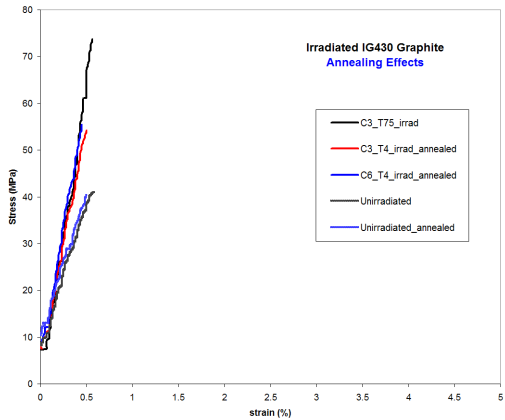

Figure 12: Stress-strain and hardening of IG-430

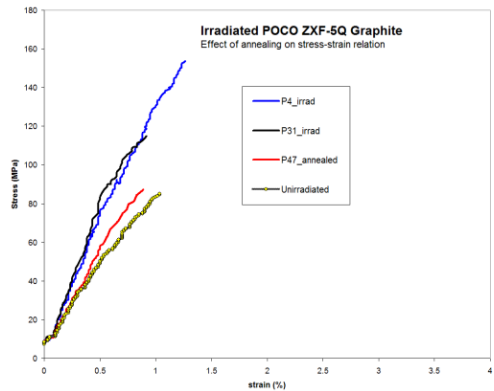

Figure 13: Stress-strain and hardening of POCO graphite

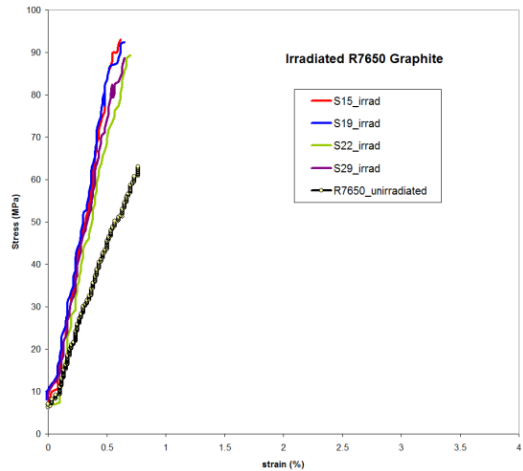

Figure 14: Stress-strain and hardening of R7650 graphite

The ongoing experimental effort confirmed that C/C and graphite structures in direct contact with water even at temperatures far below oxidizing thresholds and under energetic proton irradiation experience structural integrity degradation at fluence levels far below those safely withstood in fission reactors. Irradiation in argon environment appears to have delayed the initiation of degradation possibly due to the higher irradiating temperatures amongst other factors. From the study thus far on graphites the annealing effect of operating temperatures has been qualified and quantified while in combination with the stress-strain behavior a delineation of the performance anticipated from the different graphites studied can be deduced. Of significance was the ability to correlate and compare the observed changes in the modulus of elasticity with indirect measurements deduced from annealing/ultrasonic tests. The study also confirmed the significant strength degradation due to irradiation of h-BN attributed to the production of He and H via the (n,α), (n,p), (p,α), and (p,p) reactions. Specifically, in the case of boron there is a particularly large (n,α) cross section for boron-10 isotope, which makes up approximately 20% of natural boron.

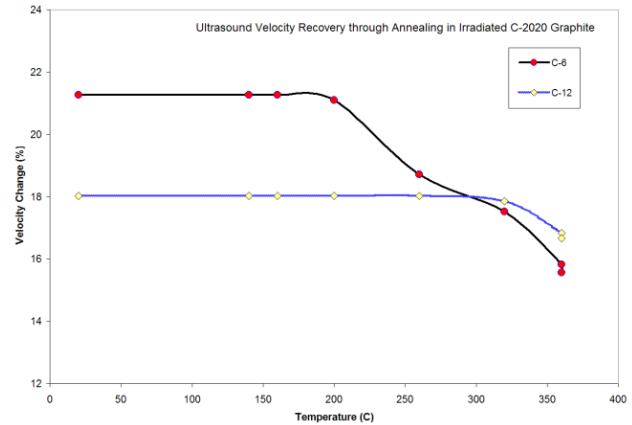

Figure 15: Sound velocity restoration through annealing in irradiated C-2020 graphite

## SUMMARY

Presented in this paper are results of an extensive irradiation study using the BNL Linear Isotope Produce facility and utilizing the 200 MeV Linac protons to irradiate four different graphite grades, h-BN and 3D C/C in water and argon atmospheres. The objectives were to address and confirm limiting proton fluences in C/C fiber composite in conjunction with the operating environment, study the performance of different graphites under consideration for target materials at LBNE under similar conditions of irradiation temperature, proton energy and fluence and, finally, explore close neighbors to graphite such as h-BN.